\documentclass[prl,twocolumn,showpacs,superscriptaddress ,preprintnumbers,amsmath,amssymb,floatfix,sort]{revtex4}

\usepackage{graphicx}
\usepackage{dcolumn}
\usepackage{bm}
\usepackage[mathscr]{eucal}
\usepackage[nooneline]{subfigure}
\newcommand{\microns}{$\mu$m}
\newcommand{\bra}[1]{\langle#1\vert}
\newcommand{\ket}[1]{\vert#1\rangle}
\newcommand{\op}[2]{\ket{#1}\bra{#2}}
\newcommand{\inner}[2]{\bra{#1} #2\rangle}

\newcommand{\eqn}[1]{Eq.~(\ref{#1})}
\newcommand{\eqns}[1]{Eqs.~(\ref{#1})}
\newcommand{\ignore}[1]{}
\newcommand{\commute}[2]{[#1,#2]}

\newcommand{\fig}[1]{Fig.~\ref{#1}}

\DeclareMathOperator{\sech}{sech}

\newcommand{\lind}{\mathcal{D}}
\newcommand{\meff}{m_\mathrm{eff}}
\newcommand{\evolution}{\mathcal{L}}

\newcommand{\sys}{\mathrm{sys}}

\newcommand{\hsys}{H_\sys}
\newcommand{\hint}{H_\mathrm{int}}
\newcommand{\htot}{H_\mathrm{tot}}

\newcommand{\hdot}{H_\mathrm{dots}}
\newcommand{\hLL}{H_\mathrm{MES}}

\newcommand{\dispersion}{\epsilon_D}
\newcommand{\annc}{c}
\newcommand{\crec}{c^\dagger}
\newcommand{\annb}{b}
\newcommand{\creb}{b^\dagger}

\newcommand{\vel}{v_0}
\newcommand{\F}{\mathcal{F}}
\newcommand{\barF}{\overline{\mathcal{F}}}
\newcommand{\rhoerror}{\Delta\rho}
\newcommand{\deltagauss}{\delta_\nu}
\newcommand{\bout}{b_\mathrm{out}}
\newcommand{\bin}{b_\mathrm{in}}

\newcommand{\chanwf}{\phi}
\newcommand{\cyclotron}{\omega_B}
\newcommand{\scattering}{\epsilon_s}
\newcommand{\spinup}{\ket{\uparrow}}
\newcommand{\spindown}{\ket{\downarrow}}
\newcommand{\gammaup}{{\gamma_\uparrow}}
\newcommand{\gammadown}{{\gamma_\downarrow}}
\newcommand{\tunprob}{\mathcal{T}^2}
\newcommand{\contactlength}{l_c}
\newcommand{\magneticlength}{l_B}
\newcommand{\muev}{$\mu$eV}

\newcommand{\mesorabi}{\Omega}
\newcommand{\mesokappa}{\kappa}

\begin{document}


\title{Mesoscopic one-way channels for quantum state transfer via the Quantum Hall Effect}

\author{T.M. Stace}
\affiliation{Cavendish Laboratory, University of Cambridge,  Madingley Road, Cambridge CB3 0HE, UK}
\affiliation{DAMTP, University of Cambridge, Wilberforce Rd, CB3 0WA, UK}
\email{tms29@cam.ac.uk}
\author{C.H.W. Barnes}
\affiliation{Cavendish Laboratory, University of Cambridge,  Madingley Road, Cambridge CB3 0HE, UK}
\author{G.J. Milburn}
\affiliation{Centre for Quantum Computer Technology, 
University of Queensland, St Lucia, QLD 4072, Australia}

\date{\today}



\begin{abstract}
We show that the one-way channel formalism of quantum optics has a physical realisation in electronic systems.  In particular, we show that magnetic edge states  form unidirectional quantum channels capable of coherently transporting electronic quantum information.  Using the equivalence between one-way photonic channels and magnetic edge states, we adapt a proposal for quantum state transfer 
 to mesoscopic systems using edge states as a quantum channel, and show that it is feasible with reasonable experimental parameters.  We discuss how this protocol may be used to transfer information encoded in number, charge or spin states of quantum dots, so it may prove useful for transferring quantum information between parts of a solid-state quantum computer. 
\end{abstract}

\maketitle
There is growing interest in using mesoscopic channels for quantum information processing tasks  \cite{bee03,sam04}, and mesoscopic analogues of \emph{one-way} channels have been considered abstractly \cite{mil00,gardiner03}. 
  One-way quantum channels  
 preserve quantum coherence, and have the additional property that forward and reverse propagating modes are distinguishable.  These channels are useful for describing the dynamics of a system coupled to a non-classical source \cite{gar00} and have been proposed for use in a quantum state transfer (QST) protocol \cite{cir97}.    The prototypical example of a one-way channel comes from quantum optics: an optical fibre with a Faraday isolator.  A magnetic field  in the Faraday isolator, which breaks time-reversal symmetry in the channel, correlates propagation direction with polarisation, making counter-propagating modes distinguishable  \cite{gar00}.   To date, mesoscopic realisations of this kind of channel have not been discussed.   
 
 Here we propose magnetic edge states \cite{hal82} of a 1D wire as a physical realisation of one-way quantum channels, and discuss their application for QST in a mesoscopic system.  In the quantum Hall effect (QHE) a magnetic field applied normal to the wire induces the formation of edge states along each edge of the wire, quantised in units of the cyclotron energy \cite{kli80,but88}.  
States on each edge propagate in a definite direction, so backscattering between counter-propagating modes is suppressed.  This accounts for the conductance plateaus observed in the QHE \cite{kli80}.
%

Bound edge states are formed by creating a local potential minimum (maximum) using surface gates to define a quantum (anti)dot.  Resonant tunnelling via \mbox{(anti)dots} coupled to extended edge states 
 has been experimentally observed \cite{vaa94,kir94,mac95}.  The tunnelling rates are tuneable using external magnetic and electric fields.

Coherent superpositions of single electrons in different edge states have been observed experimentally \cite{ji03}.  High visibility fringes ($\sim0.6$ at 20 mK) were seen in a Mach-Zender interferometer \cite{born99} consisting of edge states connected by tunnel barriers. This experiment demonstrates the possibility of transferring quantum information via edge states over distances of 10 \microns\ or more.  
 
Using edge states as one-way quantum channels, we show how to implement a proposal for QST  in a mesoscopic system. 
\citet{cir97} describe a protocol for transferring the quantum state of one two-level atom in a cavity, connected by optical fibre, to another identical system using shaped control pulses applied to each atom.  Here, we map this protocol to a system consisting of  quantum dots connected by magnetic edge states.  This may prove useful for transferring quantum information between parts of a solid-state quantum computer.  We also consider sources of error in the protocol and conclude that these are not debilitating.

\begin{figure}
\includegraphics[width=7cm]{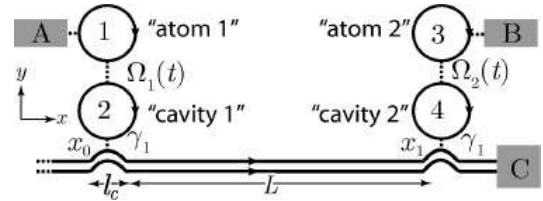}
\caption{\label{fig:schematic}Quantum dots coupled to one another and to nearby edge states.  Dotted lines indicate tunnel contact.  The grey rectangles indicate Ohmic contacts.  Dots are labelled as ``atom'' or ``cavity'' to make clear the analogy with the atom-optical scheme \cite{cir97}.}
\end{figure}

 The QHE is understood  by solving the 
Schr\"odinger equation for an electron in a lateral potential, $U(y)$, and a magnetic field, $B$ \cite{hal82}. 
The eigenmodes are of the form $\psi_{n,k}(x,y)=e^{i k x}\chi_n(y)$.  
For a linear lateral potential, $U(y)=e E y$
, the  eigenenergies are $\varepsilon_{n,k}=
\hbar\, \vel k+(n+\frac{1}{2})\hbar\cyclotron$ where $\vel=-E/B>0$, $\cyclotron=eB/\meff$ is the cyclotron frequency and $n\in{\mathbb{Z}}^+$, producing subbands of edge states.  Thus, ideally, edge states are dispersionless, $\omega=\vel k$, with group velocity $\vel$.

The system of dots in tunnel contact with magnetic edge states is shown schematically in \fig{fig:schematic}.  The system Hamiltonian  is $\htot=\hdot+\hLL+\hint$ \cite{but92}, 
\begin{eqnarray}
\frac{\hdot}{\hbar}&=&\sum_{i=1}^4\omega_i  \crec_i \annc_i+\mesorabi_1(\crec_1\annc_2+\crec_2\annc_1)+\mesorabi_2(\crec_4\annc_3+\crec_3\annc_4),\nonumber\\
\frac{\hLL}{\hbar}&=&\int d\omega g(\omega)\,\omega\,\creb(\omega)\annb(\omega),\nonumber\\
\frac{\hint}{\hbar}&=&\int d\omega (\mesokappa_1(\omega)(\creb(x_0,\omega)\annc_2+\crec_2\annb(x_0,\omega))\nonumber\\
&&\quad\quad{}+\mesokappa_2(\omega)(\creb(x_1,\omega)\annc_4+\crec_4\annb(x_1,\omega))).\nonumber
\end{eqnarray}
Here $\crec_i$ creates a dot $i$ electron, $\creb(\omega)$ creates an edge state electron with energy $\hbar\omega$, $\creb(x,\omega)=\creb(\omega)e^{i \omega x/\vel}$, $g(\omega)$ is the density of states, 
$\kappa(\omega)$ is the tunnelling rate between dots and an edge states, $x_{0,1}$  are locations of the dots and $\Omega_i(t)$ are tuneable  tunnelling  rates between the dots.  
Tunnelling rates decrease exponentially with distance from the dot,  so we only consider the nearest, resonant subband.

Non-linearities in $U(y)$ introduce dispersion.  A small quadratic term in the lateral potential, $U(y)=e E y+\frac{1}{2}\meff\omega_0^2 y^2$, results in the dispersion relation $\omega=
\vel k+\alpha k^2$, where $\alpha={\omega_0^2\hbar}/{(2\cyclotron^2 \meff)}$.  
As discussed later, for typical experimentally relevant parameters, $\alpha$ is small, and we treat this as a source of error in the protocol.

\paragraph{Input-output relations.}
The input-output formalism relates the output channel of a quantum system to its input.  
Formally, if $g(\omega)\omega$ and $\mesokappa_i(\omega)\equiv\sqrt{\gamma_i/2\pi}$ were constant over relevant frequencies around the dot energies, $\omega_i$, then the Markov approximation applies \cite{gar00}, and the input-output relation for the first system of dots is
\begin{equation}
\bout(x_{0},t)=\bin(x_{0},t)+i\sqrt{\gamma_{1}}\annc_{2}(t),\label{eqn:firstinout}
\end{equation}
 where $x_{0}$ is the location of the first pair of dots and $\annb(x,t)=\int_{-\infty}^{\infty}d\omega \,\annb(\omega)e^{i \omega(t-x/\vel)}$.  A similar relation holds for the second pair.  For a dispersionless channel, the input to the second system is given by
 \begin{equation}
 \bin(x_1,t)=\bout(x_0,t-L/v_0).\label{eqn:linearchannelio}
\end{equation}where $L=x_1-x_0$.
By virtue of the fact that back-scattering is strongly suppressed in edge states, we can neglect the effect of the second system on the first.

For a dispersing channel,  the  input to the second system is related to the output of the first by \cite{thesis}
\begin{equation}
\bin(x_1,t)=\deltagauss(t)*\bout(x_0,t-\tau)\label{eqn:dispersivechannel}
\end{equation}
where $\tau=L/\vel$, $\deltagauss(t)=(i\pi\nu)^{-1/2}e^{-t^2/i\nu}$, $\nu={{4 \alpha \tau}/{ v_0^2}}$ and the convolution $f(t)*g(t)=\int dt' f(t)g(t-t')$.

\paragraph{Number eigenstate QST.}\label{sec:QSTprimitive}
Witihin the Markov approximation, this model for quantum dots in tunnel contact with edge states parallels the atom-optical system presented in \cite{cir97}.  In particular, in the \emph{ideal} case where  the dot energies are equal ($\omega_i=\bar\omega$), the channel-dot coupling gates are equal, $\gamma_{1,2}=\gamma$, the channel is dispersionless and does not scatter, then the Hamiltonian is formally identical to that of \cite{cir97}.  

The physical correspondence to \cite{cir97} is shown in \fig{fig:schematic}.  We identify an electron on dot 1 (or 3) with an atomic excitation 
and an electron on dot 2 (or 4) with a cavity photon. 
  The computational basis for each dot is the absence, $\ket{0}_i$, or presence, $\ket{1}_i=\crec_i\ket{0}_i$ of an electron.  
  Tunnelling rates, $\Omega_i(t)$, between dots 
may be controlled via external gates. 
 Gate pulses that implement QST have been derived previously \cite{cir97,sta02}. 

Finally, we identify optical fibre modes in \cite{cir97} with edge states.  In both mesoscopic and optical systems, the presence of a magnetic field breaks time-reversal symmetry of the system so that counter-propagating modes are in principle distinguishable.  In this way, edge states are the electronic analogues of optical one-way channels, and the theory for one-way quantum channels \cite{gar00} may be applied.

Thus the QST protocol described in \cite{cir97} can be used to transfer the electronic state of dot 1 to dot 3 using edge states as a one-way quantum channel.  In this scheme the computational basis states are the eigenstates of the dot number operator, $\{\ket{0}_i,\ket{1}_i\}$.  Therefore a superposition such as $\ket{\psi}=c_0\ket{0}+c_1\ket{1}$ can be transferred coherently from dot 1 to dot 3 using this protocol, i.e. $\ket{\Psi_i}=\ket{\psi}_1\ket{0}_2\ket{0}_3\ket{0}_4\rightarrow\ket{\Psi_f}=\ket{0}_1\ket{0}_2\ket{\psi}_3\ket{0}_4$.

The dynamics of the density matrix for the system of dots is given by the master equation  \cite{sta02},
\begin{eqnarray}
\dot{\rho}&=&-i\commute{\hsys}{\rho}/\hbar+{\gamma_1} \lind[\annc_2]\rho+{\gamma_2} \lind[\annc_4]\rho\nonumber\\
&&\quad{}-\gamma\{\commute{\crec_4}{\annc_2\rho}+\commute{\rho\crec_2}{\annc_4}\}\equiv\evolution [\rho],\label{eqn:idealME}
\end{eqnarray}
where $\gamma=\sqrt{\gamma_1\gamma_2}$ and $\lind[c]\rho\equiv c\rho c^\dagger-(c^\dagger c\rho+\rho c^\dagger c)/2$.  
A simple pulse shape that effects ideal quantum state transfer is  
$
\Omega_i(t)={\gamma_i} \sech({\gamma_i}t/2)/2$ \cite{sta02},  
which we will use for evaluating various sources of error in the protocol.  For this pulse shape, the system of quantum dots remains in a pure state, given by
\begin{eqnarray}
\ket{\psi(t)}&=&c_0\ket{0000}+c_1\{\alpha(t)\ket{1000}+\alpha(-t)\ket{0010}\nonumber\\&&\quad\quad{}+{\beta(t)}(\ket{0100}-\ket{0001})/{\sqrt{2}} \},
\end{eqnarray}
$\alpha(t)=(1-\tanh(\gamma t/2))/2$ and $\beta(t)=i \sech(\gamma t/2)/\sqrt{2}$.

\begin{figure}
\subfigure[]{\includegraphics[height=2.1cm]{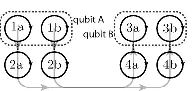}\label{fig:chargequbit}}\hfil
\subfigure[]{\includegraphics[height=2.1cm]{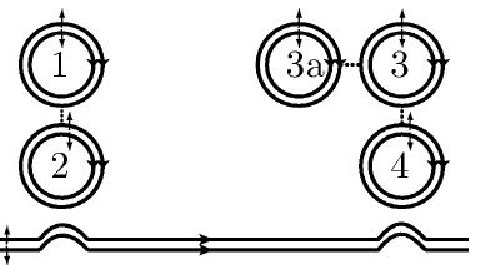}\label{fig:spinqubit}}
\caption{(a) QST for charge qubits. (b) QST for spin qubits.}
\end{figure}

\paragraph{Charge QST.}\label{sec:QSTcharge}
Encoding a qubit in the number state of electrons is impractical due to the conservation of electron number. 
 However, the scheme outlined above serves as a primitive to implement QST for a charge qubit encoded in the position of an electron in a double-well.  The  charge qubit QST protocol simply repeats the primitive protocol twice, as illustrated in \fig{fig:chargequbit}.  Qubit $A$ is encoded in the position of an electron distributed between dots 1a and 1b. 
To transfer the state of qubit $A$ to qubit $B$, the protocol first transfers the state of dot 1a to dot 3a, then the state of dot 1b to dot 3b.  The ideal protocol is unitary \cite{cir97}, so it may be concatenated as described.

\paragraph{Spin QST.}
The primitive transfer protocol described above may also be used to transfer a spin qubit from  one dot to another,  achieved using a  similar two step process as described above.  This relies on the spin dependence of the tunnelling rates \cite{mac95}, due to the different spatial configuration of spin-polarised edge states. 
 Suppose the spin-up states, $\spinup$, occupy the outer edges, as shown in \fig{fig:spinqubit}.  Since $\spinup$ states are in closer proximity to one another than $\spindown$ states, the tunnelling rate, $\gammaup$, between $\spinup$  states may be much larger than $\gammadown$. 
Therefore, the gate pulses to transfer a $\spinup$ electron from dot 1 to dot 3 are much faster than those required to transfer a $\spindown$ electron.  If an electron were in a state \mbox{$\ket{\psi}_1=c_\downarrow\spindown_1+c_\uparrow\spinup_1$}, then applying fast gate pulses, determined by $\gammaup$, would have a small effect on the $\spindown$ component of $\ket{\psi}$ for which tunnelling rates are proportional to $\gammadown$.
After this, the $\spindown$ component is transferred from dot 1 to dot 3 using pulse rates determined by $\gammadown$.  
However, turning on the coupling between dots 3 and 4 leads to a leakage error in the $\spinup$ component which was previously transferred. 
To circumvent this, the $\spinup$ component is swapped from dot 3 to an ancilla, dot 3a during the transfer of the $\spindown$ component.  Once this process is completed the $\spinup$ component is swapped back to dot 3.

\paragraph{Error sources.}
Away from the ideal case, the transfer fidelity is reduced, so we analyse the sensitivity of the protocol to small perturbations from the ideal.  For each parameter in the system, labelled generically as $p$, the transmission fidelity due to variations in $p$ is given by $\F_p=1-\barF_p$, where $\barF_p\equiv\bra{\Psi_f}\rhoerror_p\ket{\Psi_f}$ and $\rhoerror_p\equiv\op{\Psi_f}{\Psi_f}-\rho_p(\infty)$ is the error in the final state, $\rho_p(\infty)$, of the system due to errors in $p$.  
$\barF_p$ will be largest when the initial state of dot 1 is $\ket{1}_1$, 
so we use the initial state $\ket{\Psi_i}=\ket{1}_1\ket{0}_2\ket{0}_3\ket{0}_4$.  The desired final state is then $\ket{\Psi_f}=\ket{0}_1\ket{0}_2\ket{1}_3\ket{0}_4$. 
Since the errors are presumed small, they can be analysed independently, by solving  \eqn{eqn:idealME} for small perturbations of each parameter.

There is an amplitude for an electron in an edge state to be scattered into nearby states.  This is analogous to photon scattering, so we model it in the same way \cite{gar00}. 
\eqn{eqn:idealME} is modified by the replacement 
$\gamma\rightarrow \gamma\sqrt{1-\scattering}$, where $\scattering$ is the scattering probability for the channel.  The scattering probability depends on the transmission length, $L$, according to  $\scattering=1-e^{-L/L_s}$, where $L_s$ is the scattering length \cite{kom92}.
$L_s$ depends strongly on the temperature and purity of the sample, but for $T<5$ K, $L_s \gtrsim 100$ \microns\ has been reported \cite{ji03,kom92}.

{\begin{table}\setlength{\tabcolsep}{6pt}
\begin{tabular}{|c|c|c|c|c|c|}
\hline
   $p$ & 
   $\delta \omega_{1}$ &
   $\delta \omega_{2}$ &
 $\delta \gamma$    &
   $\scattering$
   &  $\dispersion$   \\ \hline
   $\barF_p$ & $4.29\, \delta \omega_1^2$ & $ 1.00\,\delta \omega_2^2$ & 
 $0.25\, \delta\gamma^2$& $1.00\,\scattering$
   & $0.02\dispersion^2$ 
   \\
\hline
\end{tabular}
  \caption[Transfer errors]{Lowest order terms in infidelity due to systematic errors. Dimensionless parameters: $\delta\omega_i={(\omega_i-\bar\omega)}/{\gamma_1}$, $\delta\gamma=1-{\gamma_2}/{\gamma_1}$,  $\scattering=1-e^{-{L}/{L_S}}$ and $\dispersion={{\alpha \tau\gamma^2}/{ v_0^2}}$.  \label{tab:errors}}
\end{table}}

Accounting for channel dispersion is more complicated, since the original formulation of one-way quantum channels implicitly assumes a dispersionless channel.  
 There are difficulties in generalising the input-output formalism to include dispersion, so we employ a different approach to estimate these errors.
After the electron has tunnelled into the channel, 
 the channel is in a superposition of zero and one electrons.  The amplitude for an electron to be at position $x$ at time $t$ is found from  \eqns{eqn:firstinout} and (\ref{eqn:linearchannelio}) to be $\bra{0}b(x,t)\ket{\chanwf}=\sqrt{\gamma/2}{\beta(t-\tau_x)}$, where $\ket{\chanwf}$ is the  channel state and 
 $\tau_x=(x-x_0)/\vel$ is the propagation time to position $x$.
Dispersion  convolves this amplitude with the channel transfer function according to \eqn{eqn:dispersivechannel}, so 
 the transmission fidelity  is then
\begin{equation}
\F=|\inner{\tilde\chanwf}{\chanwf}|^2=|\int_{-\infty}^\infty dt\, \tilde\beta^*(t)\beta(t)|^2,\label{eqn:dispersionfidelity}
\end{equation}
where $\tilde\beta=\deltagauss*\beta$. We expand  $\tilde\beta(t)$ in powers of $\nu$ then  
evaluate \eqn{eqn:dispersionfidelity} yielding $\F=1-\dispersion^2/45+...$, where $\dispersion=\nu\gamma^2/4$ is the dimensionless  dispersion strength.


Table \ref{tab:errors} lists the various dimensionless parameters in which errors may occur, along with the associated infidelities, which are computed as described earlier.  The protocol is quadratically sensitive to all parameter variations except for scattering.  Clearly, when the dots are slightly out of resonance with one another ($\delta \omega_i\neq0$) or if $\gamma_1\neq\gamma_2$, there is a corresponding error in the protocol.  

For a given error rate, we may use the tabulated results  to estimate the tolerable deviations of the parameters from ideal.  Time and  energy scales for the protocol are determined by $\gamma$, so we estimate this quantity for a typical experimental system.  
Suppose the coupling between dot 3 and the edge state channel is operated with a tunnelling probability of $\tunprob$.  This is given by $\tunprob=(h g \gamma)^2$ \cite{dat95}, where $g$ is the density of edge states in energy space.  Roughly,  $g\approx1/\Delta E$, where $\Delta E$ is the bandwidth of the tunnelling interaction given by 
\begin{equation}
\Delta E={\Delta k}\,{d E}/{d k}=({2\pi}/{\contactlength})\hbar\vel={h \vel}/{\contactlength},
\end{equation}
 where $\contactlength$ is the length over which the bound and extended states are in contact, shown in \fig{fig:schematic}.  
Therefore, 
$\tunprob=({\contactlength \gamma}/{\vel})^2\leq1
$, so  $\gamma\leq{\vel}/{\contactlength}$.  Thus,  the inverse of the time it takes an electron to pass the dot gives a bound on $\gamma$.

To estimate $\vel=-E/B$, we take $B\sim 1$ T (set by an external source), and $E$ which we estimate from the potential profile at the edge of the wire due to external gate voltages and screening effects \cite{kat00}. 
An upper estimate for $E$ is given by the slope of the lateral potential in the incompressible region \cite{kat00}.   The voltage  across the incompressible region is of order $V_i=\hbar \cyclotron/e$ and $\cyclotron=176$ GHz  for bulk GaAs in a 1 T field \cite{kat00thesis}. The width of the incompressible region is about the magnetic length, $\magneticlength=\sqrt{{\hbar}/{e B}}=25.6$ nm in a 1 T field.   Then $E_{\max}\approx V_i/ \magneticlength$ and we find that $\vel\leq67$ \mbox{km s$^{-1}$}.  
 Thus we estimate $\vel=10^4$ \mbox{m s$^{-1}$}  for the group velocity, consistent with previous results \cite{vaa94}.

The contact length is of order the dot size, so for $\contactlength\lesssim 1$ \microns\ and  $\vel=10^4$ ms$^{-1}$ we have  $\gamma\leq 100$ \muev. 
  This corresponds to an edge  perfectly transmitted into the dot region, so the dot is ill defined.  In order that the dot be well defined, the actual tunnelling rate should be some fraction of this value, so we take $\gamma\sim10$ \muev\ 
\cite{vaa94}.

From this, bounds for the precision with which the dot energies need to be controlled can be established.  If $\delta \omega_i < 0.1$, then errors arising from non-resonant dot energies will be $<1\%$.  This requires that the dot energies be controlled to around 1 \muev.  From Table \ref{tab:errors} this is also the precision with which $\gamma_i$ needs to be controlled. 

We take $\omega_0\approx 10^{11}$ Hz \cite{kom92} and $\meff=0.067 m_\mathrm{el}$ for GaAs, so transmission over $L=100$ \microns\ gives $\dispersion=0.057$.  
Using surface gates to shape the lateral potential \cite{fra98}, $\omega_0$ could be reduced: $\omega_0=10^{10}$ Hz  would give $\dispersion<10^{-3}$.  In any case, dispersive effects are negligible.

For temperatures below 1 K, $L_s\sim 1$ mm \cite{kom92}, which is much longer than the dephasing length of $L_d\sim10$ \microns\ at $T\sim 20$ mK \cite{ji03}.  The dephasing time scale is $\tau_d=L_d/\vel\sim1$ ns,  comparable with recently measured decoherence rates of charge qubits \cite{hay03}.  These results suggest that  dephasing will be a significant issue for implementing QST over distance much greater than 10 \microns.

To test our conclusions, we propose measuring currents between the Ohmic contacts shown in \fig{fig:schematic}.  Ideally, cycling the protocol at a frequency $f$ will result in a current between contacts A and B,  $I_{\mathrm AB}=e f$. Errors at a rate $p$ produce  currents  $I_{\mathrm AB}=(1-p) e f$, and  $I_{\mathrm AC}=p e f$.  Optimising $I_{\mathrm AB}$ also provides a method for tuning dot energies and tunnelling rates.


In conclusion, we have proposed magnetic edge states as physical realisations of one-way channels in mesoscopic systems.  Using edge states as quantum channels linking systems of quantum dots, we have described a mesoscopic analogue of an atom-optical system capable of implementing quantum state transfer.  This may prove useful for transferring information between parts of a solid-state quantum computer.  
Our proposal builds on recent experimental results demonstrating interference in an edge state interferometer, and our error analysis indicates that the protocol is experimentally feasible.  

TMS acknowledges the Hackett committee and the CVCP for financial support.  CHWB acknowledges the EPSRC for funding.  GJM acknowledges the support of the ARC through the IREX grant.  We would like to thank Sean Barrett, Hsi-Sheng Goan and Andre Luiten for useful conversations.


\end{document}